\documentstyle[aps,pre,twocolumn]{revtex}
\begin{document}

\title{Dynamical regimes in DPD}
\author{Pep Espa\~{n}ol and Mar Serrano}
\address{Departamento de F\'{\i}sica Fundamental, Universidad Nacional \\
de Educaci\'on a Distancia, C/ Senda del Rey s/n, E-28040, Madrid,\\
Spain}
\date{\today}
\maketitle

\begin{abstract}
We discuss theoretically the behavior of the velocity autocorrelation
function in the Dissipative Particle Dynamics model. Two dynamical
regimes are identified depending on the dimensionless model
parameters. For low frictions a mean field behavior is observed in
which the kinetic theory for DPD provides good predictions. For high
frictions, collective hydrodynamic effects are dominant. We have
performed numerical simulations that validate the theory presented.
\end{abstract}

\section{Introduction}

The Dissipative Particle Dynamics model allows to simulate
hydrodynamics at mesoscopic scales in which thermal fluctuations are
important \cite{hoo92,esp95}. For this reason, it appears as a
good simulation technique for the study of complex fluids like polymer
or colloidal suspensions where both hydrodynamic interactions and
Brownian motion are important
\cite{sch95,boe97,cov97,gro97,ijmp97}. Being an off lattice technique,
it does not suffer from the restrictions imposed by the lattice as in
Lattice Gas Cellular Automata or the Lattice Boltzmann approach and it
is much more flexible for modeling.

Even though the technique has a well sounded theoretical support and
many applications have been undertaken, there is at present no
systematic study of the region of parameters much suitable for
simulation of particular hydrodynamic problems. In addition, recent
simulations \cite{pag98} have shown deviations from the transport
coefficients predicted by the kinetic theory developed by Marsh {\em
et al.} \cite{mar97}. The two approximations involved in this kinetic
theory are the small gradient expansion around local equilibrium and
the molecular chaos hypothesis. However, it is difficult to
investigate the origin of the discrepancies between theory and
simulations within the kinetic theory context. The theory just produces
the explicit expressions for the transport coefficients with no hint
about its range of validity. It has been suggested that it is precisely
in the region of parameters where kinetic theory fails where it is more 
sensible to conduct simulations that reproduce hydrodynamic behavior 
\cite{pag98}.

We shed some light into the problem by presenting a theory that allows
to compute the velocity autocorrelation function (vaf) of the
dissipative particles. The theory is based on the physical picture in
which the DPD particles are regarded as Brownian-like particles moving
in an environment created by the rest of DPD particles. Strictly
speaking, though, they are not Brownian particles because the total
moment of the system is conserved. This approach was introduced by
Groot and Warren as a way of computing the diffusion coefficient
\cite{gro97}. We identify the basic dimensionless parameters which
allow to classify and discuss the dynamical regimes displayed by the
model. By assuming that the environment of DPD particle behaves
hydrodynamically, it is possible to obtain an explicit analytical
expression for the velocity autocorrelation function. Finally, we
present numerical simulations that allow to validate the hypothesis
made in the theory.

\section{The DPD model}

The stochastic differential equations (SDE) that govern the position
${\bf r}_{i}$ and velocity ${\bf v}_{i}$ of the $i$-th 
particle of mass $m$ in DPD are given by \cite{esp95}

\begin{eqnarray}
d{\bf r}_i&=&{\bf v}_idt,
\nonumber \\
md{\bf v}_i
&=&-\gamma m\sum_{j}\omega(r_{ij})
({\bf e}_{ij}\!\cdot\!{\bf v}_{ij})
{\bf e}_{ij}dt 
\nonumber\\
&+&\sigma\sum_j\omega^{1/2}(r_{ij}){\bf e}_{ij}dW_{ij}, 
\label{sde}
\end{eqnarray}
where the following quantities are defined 
\begin{eqnarray}
{\bf e}_{ij}
&\equiv&\frac{{\bf r}_{ij}}{r_{ij}}
\nonumber \\
{\bf r}_{ij}&\equiv&{\bf r}_i-{\bf r}_j,
\nonumber \\
r_{ij}&\equiv&|{\bf r}_i-{\bf r}_j|,
\nonumber \\
{\bf v}_{ij}
&\equiv& {\bf v}_i-{\bf v}_j.
\label{def1}
\end{eqnarray}
In order to compare with the kinetic theory in Ref. \cite{mar97}, it is
assumed that the usual conservative force is not present. The noise
amplitude $\sigma$ is given by the detailed balance condition

\begin{equation}
\sigma=(2k_B T\gamma m)^{1/2},
\label{fd}
\end{equation}
where $T$ is the temperature of the equilibrium state towards which
the system relax (if the boundary conditions allow for it) and $k_B$
is Boltzmann's constant. Finally, $dW_{ij}=dW_{ji}$ are independent
increments of the Wiener process that obey the It\^o calculus rule
\begin{equation}
dW_{ij}dW_{{i'}{j'}}
=(\delta_{i{i'}}\delta_{j{j'}}
+\delta_{i{j'}}\delta_{j{i'}})dt,
\label{ito}
\end{equation}
i.e., $dW_{ij}$ is an infinitesimal of order $1/2$ \cite{gar83}.  The
dimensionless weight function $\omega(r)$ is normalized according to
\cite{hoo92}
\begin{equation}
\int d{\bf r} \omega(r) = \frac{1}{n},
\label{norm}
\end{equation}
where $n$ is the number density of the system. In this paper we will work
in two spatial dimensions (2D) and select the following weight function 
with range $r_c$,

\begin{equation}
\omega(r) =\frac{3}{\pi r_c^2 n}\left(1-\frac{r}{r_c}\right),
\label{ome}
\end{equation}
if $r<r_c$ and zero if $r>r_c$. 

We discuss now which are the fundamental parameters for the DPD
model. By appropriate choice of units of mass, time and space it is
always possible to reduce the number of relevant parameters of the
model. It is obvious that the dynamical regimes are independent of the
units used, and will depend on {\em dimensionless} parameters
only. There are six parameters in the model: $m, \gamma, r_c, k_BT,
\lambda, L$, where $\lambda$ is the average distance between
particles, related to the number density $n$ of particles by $\lambda
= n^{-1/d}$, $d$ is the space dimension, and $L$ is the box size (or any
other boundary length scale). From
these six parameters we can form three dimensionless parameters. By
defining the thermal velocity $v_T=(k_BT/m)^{1/2}$ we select

\begin{eqnarray}
\Omega &\equiv &\frac{\gamma r_c}{d v_T}=\frac{\tau_T}{d \tau_\gamma},
\nonumber \\
s &\equiv& \frac{r_c}{\lambda} ,
\nonumber\\
\mu &\equiv&\frac{L}{r_c}.
\label{dim1} 
\end{eqnarray}
The physical meaning of these parameters is as follows: $\tau_T$ is
the time taken by a particle moving at the thermal velocity to move a
distance $r_c$, whereas $\tau_\gamma=\gamma^{-1}$ is the time
associated to the friction. Therefore, the {\em dimensionless
friction} $\Omega$ is the ratio of these two time scales. On the other
hand, $s$ is the {\em overlapping} between particles which is related
to the number of particles that are within the range of interaction
(the {\em action sphere}) of a given one. Finally, $\mu$ is the
dimensionless box length. These dimensionless parameters $\Omega,
s,\mu$ fix the dynamical regimes of the model.

\section{Velocity autocorrelation function}
The velocity equation in (\ref{sde}) can be written in
the form

\begin{equation}
\dot{{\bf v}}_i = -\gamma \left[\sum_{j\neq i}\omega(r_{ij})
{\bf e}_{ij}{\bf e}_{ij}\right]\!\cdot\!{\bf v}_{i}
+\frac{\gamma}{d}{\bf V}_i(t)
+\frac{\tilde{{\bf F}}_i}{m},
\label{rew}
\end{equation}
where the random force is $\tilde{{\bf F}}_i
dt=\sigma\sum_j\omega^{1/2}(r_{ij}){\bf e}_{ij}dW_{ij}$. 
We have introduced in (\ref{rew}) the {\em environment} velocity by

\begin{equation}
{\bf V}_i(t)=
d\sum_{j\neq i}\omega(r_{ij}){\bf e}_{ij}{\bf e}_{ij}\!\cdot\!{\bf v}_j.
\label{vhidro}
\end{equation}
This velocity is a weighted average of the velocities of the neighboring
particles of particle $i$. Next, we observe that the factor of ${\bf
v}_i$ in the right hand side of (\ref{rew}) can be written as

\begin{equation}
\sum_{j\ne i}\omega(r_{ij}){\bf e}_{ij}{\bf e}_{ij}
=
 \int d{\bf r}
\omega({\bf r}_i-{\bf r})
 \frac{{\bf r}_i-{\bf r}}{|{\bf r}_i-{\bf r}|}
 \frac{{\bf r}_i-{\bf r}}{|{\bf r}_i-{\bf r}|}
n({\bf r},t),
\label{mic1}
\end{equation}
where we have introduced the microscopic density field $n({\bf r},t)=
\sum_{j\neq i}\delta({\bf r}_j-{\bf r})$.  If we assume that the density field
is constant with value $n$ (which will be confirmed by the results
obtained later) then we may approximate

\begin{eqnarray}
&& \int d{\bf r}
\omega({\bf r}_i-{\bf r})
 \frac{{\bf r}_i-{\bf r}}{|{\bf r}_i-{\bf r}|}
 \frac{{\bf r}_i-{\bf r}}{|{\bf r}_i-{\bf r}|}
n({\bf r},t)
\nonumber\\
&&\approx n \int d{\bf r}
\omega({\bf r}) \frac{{\bf r}}{|{\bf r}|}\frac{{\bf r}}{|{\bf r}|}
=\frac{{\bf 1}}{d}
\label{mic2}
\end{eqnarray}
The last equality is obtained by noting that the integral is an
isotropic second order tensor, which must be proportional to the
identity (the constant of proportionality can be obtained by taking
the trace of the integral and using the normalization (\ref{norm})).

After using Eqns. (\ref{mic1}) and (\ref{mic2}) in Eqn. (\ref{rew}) one obtains
\begin{eqnarray}
d{\bf r}_i&=&{\bf v}_idt,
\nonumber \\
d{\bf v}_i
&=&-\frac{\gamma}{d}\left[{\bf v}_{i}-{\bf V}_i\right]dt 
+ \frac{\tilde{{\bf F}}_i}{m} dt.
\label{sde-mf}
\end{eqnarray}
We observe that the DPD particles behave similarly to Brownian
particles but in a systematic velocity field determined by the rest of
its neigbouring particles. The stochastic properties of the random
force are not exactly those of a Brownian particle because the total
momentum of the system is conserved but for the rest of the
development they are irrelevant.

The formal solution of Eqn. (\ref{sde-mf}) is

\begin{eqnarray}
{\bf v}_i(t)&=&\exp\{-\gamma t/d\}{\bf v}_i(0)
\nonumber\\
&+&
\int_0^t dt'\exp\{-\gamma (t-t')/d\}
\left[\frac{\gamma}{d}{\bf V}_i(t')+\frac{\tilde{{\bf F}}_i(t')}{m}\right].
\nonumber\\
\label{mfsol}
\end{eqnarray}
By multiplying this equation by ${\bf v}_i(0)$ and ${\bf V}_i(0)$ and
averaging one obtains a set of equations for the velocity
autocorrelation function at equilibrium

\begin{eqnarray}
&&\frac{1}{d}\langle {\bf v}_i(t)\!\cdot\!{\bf v}_i(0)\rangle
=\exp\{-\gamma t/d\}\frac{k_BT}{m}
\nonumber\\
&+&\frac{\gamma}{d}\int_0^t dt'\exp\{-\gamma (t-t')/d\}
\frac{1}{d}\langle{\bf V}_i(t')\!\cdot\!{\bf v}_i(0)\rangle,
\nonumber\\
&&\frac{1}{d}\langle {\bf v}_i(t)\!\cdot\!{\bf V}_i(0)\rangle =
\nonumber\\
&&\frac{\gamma}{d}\int_0^t dt'\exp\{-\gamma (t-t')/d\}
\frac{1}{d}\langle{\bf V}_i(t')\!\cdot\!{\bf V}_i(0)\rangle,
\nonumber\\
\label{acvf}
\end{eqnarray}
where use has been made of the fact that the random force is not
correlated with the velocity at present and past times  and the
property $\langle{\bf V}_i(0)\!\cdot\!{\bf v}_i(0)\rangle=0$ (which
can be checked from the definition (\ref{vhidro})). Substitution of
the second equation in (\ref{acvf}) into the first one leads to an
expression that relates the particle vaf with the
environment vaf, this is
\begin{eqnarray}
&&\frac{1}{d}\langle {\bf v}_i(t)\!\cdot\!{\bf v}_i(0)\rangle
=\exp\{-\gamma t/d\}\frac{k_BT}{m}
\nonumber\\
&&+\left(\frac{\gamma}{d}\right)^2
\int_0^t dt'(t-t')\exp\{-\gamma (t-t')/d\}
\frac{1}{d}\langle{\bf V}_i(t')\!\cdot\!{\bf V}_i(0)\rangle.
\nonumber\\
\label{acvf01}
\end{eqnarray}
The second term in the right hand side represents {\em collective
effects}. When this term is negligible we say that a {\em mean field}
approximation is valid, in which the velocity autocorrelation function
decays exponentially. The reason for the name ``mean field'' comes form
the observation that Eqn. (\ref{sde-mf}), in which the average value
$\langle {\bf V}_i\rangle=0$ is used instead of the instantaneous
value ${\bf V}_i$, produces an exponential decay of the velocity
autocorrelation function. In the Appendix of Ref. \cite{gro97} it
was computed the velocity autocorrelation function and the diffusion
coefficient of the DPD particles by using this mean field approximation.

The solution (\ref{acvf01}) is still formal because we do not know
explicitly the form of the correlation of the environment velocity
(which will be given in the next section).
Nevertheless, it is possible to extract useful information from this
expression. This is most conveniently done by taking dimensionless
variables. Let ${\overline t}$ be the dimensionless time $t v_T/r_c$,
this is, the time expressed in units in which $r_c=1$ and $v_T=1$, and
${\overline {\bf v}}={\bf v}/v_T$ a  dimensionless velocity. In these
units, Eqns. (\ref{acvf}) take the form

\begin{eqnarray}
\frac{1}{d}\langle {\overline {\bf v}}_i({\overline t})\!\cdot\!{\overline {\bf
v}}_i(0)\rangle &=&\exp\{-\Omega {\overline t}\}
\nonumber\\
&+&\Omega\int_0^{\overline t} d{\overline t}'\exp\{-\Omega ({\overline
t}-{\overline t}')\} \frac{1}{d}\langle{\overline {\bf
V}}_i({\overline t}')\!\cdot\!{\overline {\bf v}}_i(0)\rangle ,
\nonumber\\
\frac{1}{d}\langle {\overline {\bf v}}_i({\overline t})\!\cdot\!{\overline {\bf
V}}_i(0)\rangle &=& \Omega\int_0^{\overline t} d{\overline
t}'\exp\{-\Omega ({\overline t}-{\overline t}')\}
\frac{1}{d}\langle{\overline {\bf V}}_i({\overline t}')\!\cdot\!{\overline {\bf
V}}_i(0)\rangle,
\nonumber\\
\label{acvfbis}
\end{eqnarray}
and Eqn. (\ref{acvf01}) takes the form
\begin{eqnarray}
&&\frac{1}{d}
\langle {\overline {\bf v}}_i({\overline t})\!\cdot\!{\overline {\bf v}}_i(0)\rangle
=\exp\{-\Omega {\overline t}\}
\nonumber\\
&&+\Omega^2
\int_0^{\overline t}  d{\overline t}'({\overline t}-{\overline t}')
\exp\{-\Omega 
({\overline t}-{\overline t}')\}
\frac{1}{d}
\langle{\overline {\bf V}}_i({\overline t}')\!\cdot\!{\overline {\bf V}}_i(0)\rangle
\label{acvf02}
\end{eqnarray}
For later notational convenience we introduce

\begin{eqnarray}
c({\overline t})&\equiv&\frac{1}{d}
\langle {\overline {\bf v}}_i({\overline t})
\!\cdot\!{\overline {\bf v}}_i(0)\rangle,
\nonumber\\
C({\overline t})&\equiv&
\frac{1}{d}\langle{\overline {\bf V}}_i({\overline t})
\!\cdot\!{\overline {\bf V}}_i(0)\rangle.
\label{dein}
\end{eqnarray}
In order to complete the picture we need to evaluate the overall
magnitude of the correlation of the environment velocity. This is
determined by the value at the origin which is computed in the
Appendix I with the result

\begin{equation}
\frac{1}{d}\langle{\overline {\bf V}}_i(0)\!\cdot\!{\overline {\bf
V}}_i(0)\rangle =\frac{1}{s^2}\frac{3 d}{2\pi}.
\label{evt0}
\end{equation}
We observe that the magnitude of this correlation decreases with the
overlapping coefficient. This is physically meaningful because the
environment velocity is a weighted average of the velocities of the
particles that are within an action sphere. These velocities are
distributed at random and, therefore, if there are many particles
within an action sphere, the average will be proportionally smaller.

Now, several qualitative predictions concerning the different
dynamical regimes can be made from expression (\ref{acvfbis}) or
(\ref{acvf02}). For fixed $\Omega$, the large overlapping $s$ limit
produces a small contribution from the collective part and the
velocity correlation function decays in an exponential way. For fixed
overlapping $s$, when $\Omega$ is small (in the limit of small
friction or high temperature) the behavior of the vaf is again
exponential. In the opposite regime of large $\Omega$, the exponential
contribution decays in a very short time and the main contribution for
times larger than $\Omega^{-1}$ is given by the collective
term. Actually, in the limit $\Omega\rightarrow\infty$ the exponential
memory function acts as a delta function and for times larger than
$\Omega ^{-1}$ one obtains

\begin{equation}
\langle{\overline {\bf v}}_i({\overline t})\!\cdot\!{\overline {\bf v}}_i(0)\rangle
\approx
\langle{\overline {\bf V}}_i({\overline t})\!\cdot\!{\overline {\bf v}}_i(0)\rangle
\approx
\langle{\overline {\bf V}}_i({\overline t})\!\cdot\!{\overline {\bf V}}_i(0)\rangle.
\label{highomeg}
\end{equation}
The physical meaning of these expressions (\ref{highomeg}) is also
clear. When the friction is high, in a very short time the velocity of
a given particle is slaved by the average velocity of its environment.

\section{Hydrodynamic hypothesis}
The environment velocity ${\bf V}_i$ defined in  Eqn. (\ref{vhidro})
can be rewritten as an average over the action
sphere of the microscopic velocity field, this is,

\begin{equation}
{\bf V}_i(t)=d\int d{\bf r}
\omega({\bf r}_i-{\bf r})
 \frac{{\bf r}_i-{\bf r}}{|{\bf r}_i-{\bf r}|}
 \frac{{\bf r}_i-{\bf r}}{|{\bf r}_i-{\bf r}|}
n({\bf r},t){\bf v}({\bf r},t),
\label{hh1}
\end{equation}
where
\begin{equation}
n({\bf r},t){\bf v}({\bf r},t)
=\sum_{j\neq i}{\bf v}_j\delta({\bf r}_j-{\bf r}).
\end{equation}
The velocity field ${\bf v}({\bf r},t)$ obeys the
equations of hydrodynamics when its characteristic length scale is
much larger than the interparticle distance. We expect that the
average involved in ${\bf V}_i$ will be dominated by the hydrodynamic
modes whenever the range of interaction $r_c$ is much larger than the
interparticle distance $\lambda$ (i.e., large overlapping $s$). In
this section we present a calculation of the correlation function of
the environment velocity which is based on this hydrodynamic argument.

From Eqn. (\ref{hh1}), the environment velocity is
conveniently expressed in terms of the Fourier components of the
velocity field ${\bf v}({\bf k},t)$, this is

\begin{equation}
{\bf V}_i(t) = dn\int \frac{d{\bf k}}{(2 \pi)^2}
{\bf \omega}({\bf k})\!\cdot\! {\bf v}({\bf k},t) \exp\{i {\bf k}{\bf r}_i\},
\label{ve1}
\end{equation}
where we have introduced the second order tensor

\begin{equation}
{\bf \omega}({\bf k})= \int d{\bf r}\omega(r){\bf \hat{r}}{\bf \hat{r}}
\exp\{-i {\bf k }{\bf r}\}.
\end{equation}
The explicit form of this tensor when the weight function is given by Eqn. (\ref{ome})
is given in the Appendix II. The environment velocity correlation
function is given by

\begin{eqnarray}
\langle {\bf V}_i(0)\!\cdot\!{\bf V}_i(t)\rangle
&=&(d n )^2 \int \frac{ d{\bf k}}{(2 \pi)^2}\frac{ d{\bf k'}}{(2 \pi)^2}
{\bf \omega}({\bf k}){\bf \omega}({\bf k}')
\nonumber\\
&\times &
\langle  \exp\{i ({\bf k }{\bf r}_i(0)+{\bf k' }{\bf r}_i(t))\}  
{\bf v}({\bf k},0){\bf v}({\bf k'},t)\rangle.
\nonumber\\
\label{evacf}
\end{eqnarray}
We will assume that the position of particle $i$ is weakly correlated
with the velocity field ${\bf v}({\bf k},t)$ in such a way that we
can approximate

\begin{eqnarray}
&&\langle \exp\{i ({\bf k }{\bf r}_i(0)+{\bf k' }{\bf r}_i(t))\} 
{\bf v}({\bf k},0){\bf v}({\bf k'},t)\rangle 
\nonumber\\
&&\approx 
\langle \exp\{i ({\bf k }{\bf
r}_i(0)+{\bf k' }{\bf r}_i(t))\}\rangle 
\;\;\langle {\bf v}({\bf k},0){\bf v}({\bf k'},t)\rangle.
\label{2vm}
\end{eqnarray}
We further assume that the correlation of the velocity field
is given by the linear hydrodynamics result \cite{lin}

\begin{eqnarray}
&&\langle {\bf v}({\bf k'},0){\bf v}^{T}({\bf k},t)\rangle =
\frac{k_BT}{nm} (2 \pi)^{2} \delta ({\bf k}+{\bf k}')
\nonumber\\
&&\times [\exp\{-\nu k^{2}
t\} ({\bf 1} -{\bf\hat{k}} {\bf\hat{k}}) +\exp\{-\Gamma k^{2} t\} \cos{kct}
{\bf \hat{k}} {\bf \hat{k}}].
\label{vfacf}
\end{eqnarray}
Here, $\nu$ is the kinematic viscosity, $\Gamma$ is the sound absorption
coefficient, and $c$ is the sound speed of the DPD fluid. The
correlation function (\ref{vfacf}) is different from zero only when
${\bf k}=-{\bf k}'$. The first average in the right hand side of
(\ref{2vm}) is, therefore, given by the incoherent intermediate
scattering function $F_s({\bf k },t)= \langle \exp\{i {\bf k }\!\cdot\!({\bf
r}_i(0)-{\bf r}_i(t))\}\rangle$ \cite{lin}. By further assuming a
hydrodynamic behavior for this function we obtain \cite{lin}
\begin{equation}
 F_s({\bf k },t)  = \exp\{- D k^2 t\},
\label{diff}
\end{equation}
where $D$ is the self-diffusion coefficient of the DPD particles. 

The final hydrodynamic expression for the environment velocity
correlation function is found by substitution of
(\ref{2vm}),(\ref{vfacf}),(\ref{diff}), into (\ref{evacf})
\begin{eqnarray}
\frac{1}{d}\langle {\bf V}_i(0)\!\cdot\!{\bf V}_i(t)\rangle 
&=&
\frac{3d k_BT}{4\pi r_c^2 n m}
\left[\Phi\left (\frac{(\nu+D)t}{r_c^2}\right )\right.
\nonumber\\
&+&\left.
\Psi\left(\frac{(\Gamma+D)t}{r_c^2},\frac{ct}{r_c}\right )\right],
\label{hp1}
\end{eqnarray}
where the following functions are defined
\begin{eqnarray}
\Phi \left (x \right )
&=&\frac{\int d{\bf k} a^2(k) 
\exp\{-x k^{2} \}}{\int d{\bf k} a^2(k)}
\nonumber\\
\Psi \left (y,z\right )&=& \frac{\int d{\bf k}
(a(k)+b(k))^2 \exp\{-y k^{2}\} \cos{kz}}
{\int d{\bf k} (a(k)+b(k))^2} ,
\label{hp0}
\end{eqnarray}
which satisfy $\Phi(0)=1,\Psi(0,0)=1$, with the definitions for $a(k),
b(k)$ given in Appendix II.

In this way, the environment velocity correlation function is
explicitly given in terms of hydrodynamic fluid properties, i.e., the
transport coefficients of the DPD fluid. The prediction is not
complete until particular values for these transport coefficients are
provided. One possibility is to measure these transport coefficients
in a simulation.  Another possibility, which is the one we follow
here, is to use the values for $\nu, D, \Gamma, c$ provided by kinetic
theory \cite{mar97}. This has the advantage that one knows the
explicit dependence of the transport coefficients in terms of the
dimensionless parameters $\Omega,s$. Even though for some dynamical
regimes the kinetic theory predictions are not in exact quantitative
agreement with the measured transport coefficients \cite{pag98,rev98}
the discrepancies between theoretical and simulation transport
coefficients are small.

The kinetic theory results are \cite{mar97},\cite{esp98}

\begin{eqnarray}
D&=&\frac{d}{\gamma}\frac{k_BT}{m},
\nonumber\\
\nu &=&\frac{1}{2}\left[ \gamma n \frac{1}{d(d+2)}\int r^{2} \omega(r)d{\bf r}
+c^{2} \frac{d}{\gamma  n \int \omega (r) d {\bf r} }\right],
\nonumber \\
\nu_b &=& \gamma n \frac{1}{2d^{2}}\int r^{2} \omega(r)d{\bf r}
+c^{2} \frac{1}{\gamma  n \int \omega (r) d {\bf r} },
\nonumber \\
c &=&\sqrt{\frac{k_B T}{m}}.
\label{trans1}
\end{eqnarray}
The sound absorption coefficient is given by $\Gamma = 2 \nu +
\frac{1}{2}\nu_b$ for the case that the equation of state is that of
the ideal gas \cite{lin}. 

\begin{figure}[ht]
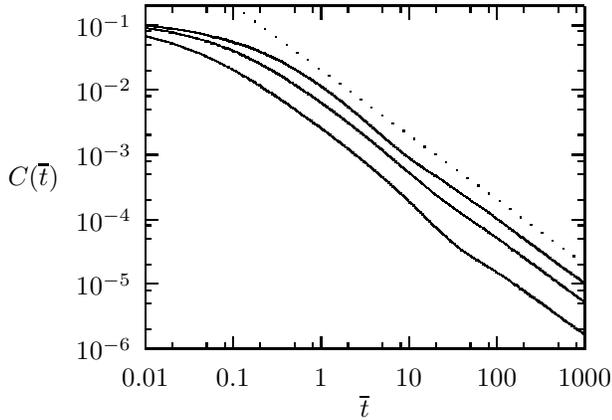
 \begin{center} 
\setlength{\unitlength}{0.240900pt}
\ifx\plotpoint\undefined\newsavebox{\plotpoint}\fi

\label{fig1}
\caption{Theoretical hydrodynamic prediction for the environment
velocity correlation function $C({\overline t})$ for a
value of $s=2.82$ and three different values for $\Omega$. Lowest
curve is for $\Omega=25$, middle curve is for $\Omega=8.3$, and upper
curve is for $\Omega =0.5$. The algebraic long time tail behavior
appears for very long times, typically when the correlation has
decayed three orders of magnitude from its initial value. Dotted line
is a curve proportional to $t^{-1}$ to guide the eye.}  
\end{center}
\end{figure}

We are now in position to write the hydrodynamic prediction (\ref{hp1}) in
terms of dimensionless variables. Substitution of (\ref{trans1}) with the
expression (\ref{ome}) into (\ref{hp1}) leads to 

\begin{eqnarray}
C({\overline t})
&=&
\frac{3 d }{4 \pi s^2 }\left[\Phi\left(\left[\frac{3}{80}\Omega +\frac{3}{2\Omega}
\right]{\overline t}\right )\right.
\nonumber\\
&+&\left.\Psi\left(
\left[\frac{9}{80}\Omega +\frac{11}{4\Omega}\right]{\overline t},{\overline t}
\right)\right].
\label{hp3}
\end{eqnarray}
We observe that the time scale of the environment velocity correlation
function is determined by $\Omega$, while its amplitude is determined
by $s$. In Fig. \ref{fig1} we show the theoretical prediction
(\ref{hp3}) for a particular value of $s$ and three different values
of $\Omega$. An algebraic dependence as $t^{-1}$ is observed for very
long times which is the celebrated long time tail $t^{-d/2}$ arising
from the diffusive character of the shear mode.

\section{Simulation results}
We have simulated Eqns. (\ref{sde}) in two spatial dimensions in a
system with periodic boundary conditions. The velocity autocorrelation
function of the DPD particles and also the environment velocity
autocorrelation function have been computed at equilibrium.

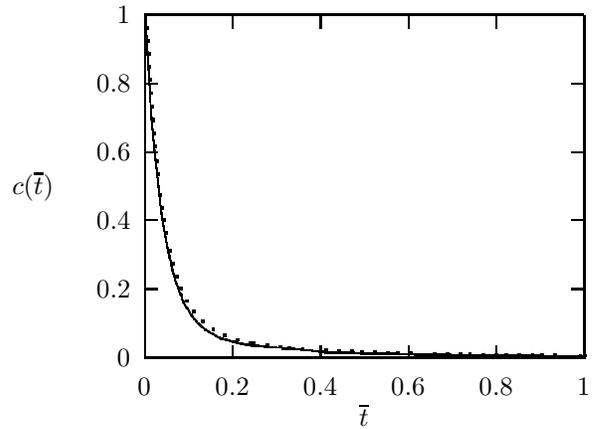
\begin{figure}[ht] 
\begin{center} 
\setlength{\unitlength}{0.240900pt}
\ifx\plotpoint\undefined\newsavebox{\plotpoint}\fi
\sbox{\plotpoint}{\rule[-0.200pt]{0.400pt}{0.400pt}}%
\begin{picture}(974,675)(0,0)
\font\gnuplot=cmr10 at 10pt
\gnuplot
\sbox{\plotpoint}{\rule[-0.200pt]{0.400pt}{0.400pt}}%
\put(220.0,113.0){\rule[-0.200pt]{166.221pt}{0.400pt}}
\put(220.0,113.0){\rule[-0.200pt]{0.400pt}{129.845pt}}
\put(220.0,113.0){\rule[-0.200pt]{4.818pt}{0.400pt}}
\put(198,113){\makebox(0,0)[r]{0}}
\put(890.0,113.0){\rule[-0.200pt]{4.818pt}{0.400pt}}
\put(220.0,221.0){\rule[-0.200pt]{4.818pt}{0.400pt}}
\put(198,221){\makebox(0,0)[r]{0.2}}
\put(890.0,221.0){\rule[-0.200pt]{4.818pt}{0.400pt}}
\put(220.0,329.0){\rule[-0.200pt]{4.818pt}{0.400pt}}
\put(198,329){\makebox(0,0)[r]{0.4}}
\put(890.0,329.0){\rule[-0.200pt]{4.818pt}{0.400pt}}
\put(220.0,436.0){\rule[-0.200pt]{4.818pt}{0.400pt}}
\put(198,436){\makebox(0,0)[r]{0.6}}
\put(890.0,436.0){\rule[-0.200pt]{4.818pt}{0.400pt}}
\put(220.0,544.0){\rule[-0.200pt]{4.818pt}{0.400pt}}
\put(198,544){\makebox(0,0)[r]{0.8}}
\put(890.0,544.0){\rule[-0.200pt]{4.818pt}{0.400pt}}
\put(220.0,652.0){\rule[-0.200pt]{4.818pt}{0.400pt}}
\put(198,652){\makebox(0,0)[r]{1}}
\put(890.0,652.0){\rule[-0.200pt]{4.818pt}{0.400pt}}
\put(220.0,113.0){\rule[-0.200pt]{0.400pt}{4.818pt}}
\put(220,68){\makebox(0,0){0}}
\put(220.0,632.0){\rule[-0.200pt]{0.400pt}{4.818pt}}
\put(358.0,113.0){\rule[-0.200pt]{0.400pt}{4.818pt}}
\put(358,68){\makebox(0,0){0.2}}
\put(358.0,632.0){\rule[-0.200pt]{0.400pt}{4.818pt}}
\put(496.0,113.0){\rule[-0.200pt]{0.400pt}{4.818pt}}
\put(496,68){\makebox(0,0){0.4}}
\put(496.0,632.0){\rule[-0.200pt]{0.400pt}{4.818pt}}
\put(634.0,113.0){\rule[-0.200pt]{0.400pt}{4.818pt}}
\put(634,68){\makebox(0,0){0.6}}
\put(634.0,632.0){\rule[-0.200pt]{0.400pt}{4.818pt}}
\put(772.0,113.0){\rule[-0.200pt]{0.400pt}{4.818pt}}
\put(772,68){\makebox(0,0){0.8}}
\put(772.0,632.0){\rule[-0.200pt]{0.400pt}{4.818pt}}
\put(910.0,113.0){\rule[-0.200pt]{0.400pt}{4.818pt}}
\put(910,68){\makebox(0,0){1}}
\put(910.0,632.0){\rule[-0.200pt]{0.400pt}{4.818pt}}
\put(220.0,113.0){\rule[-0.200pt]{166.221pt}{0.400pt}}
\put(910.0,113.0){\rule[-0.200pt]{0.400pt}{129.845pt}}
\put(220.0,652.0){\rule[-0.200pt]{166.221pt}{0.400pt}}
\put(45,382){\makebox(0,0){$c({\overline t})$}}
\put(565,23){\makebox(0,0){${\overline t}$}}
\put(220.0,113.0){\rule[-0.200pt]{0.400pt}{129.845pt}}
\sbox{\plotpoint}{\rule[-0.500pt]{1.000pt}{1.000pt}}%
\multiput(221,652)(1.263,-20.717){4}{\usebox{\plotpoint}}
\multiput(226,570)(1.233,-20.719){5}{\usebox{\plotpoint}}
\multiput(231,486)(1.908,-20.668){3}{\usebox{\plotpoint}}
\multiput(237,421)(2.379,-20.619){2}{\usebox{\plotpoint}}
\multiput(243,369)(2.574,-20.595){2}{\usebox{\plotpoint}}
\multiput(248,329)(3.825,-20.400){2}{\usebox{\plotpoint}}
\put(257.78,280.62){\usebox{\plotpoint}}
\put(262.52,260.42){\usebox{\plotpoint}}
\put(268.34,240.53){\usebox{\plotpoint}}
\put(275.74,221.14){\usebox{\plotpoint}}
\multiput(277,218)(7.983,-19.159){0}{\usebox{\plotpoint}}
\put(284.25,202.25){\usebox{\plotpoint}}
\multiput(288,196)(12.453,-16.604){0}{\usebox{\plotpoint}}
\put(295.84,185.06){\usebox{\plotpoint}}
\multiput(299,180)(14.676,-14.676){0}{\usebox{\plotpoint}}
\put(309.84,169.96){\usebox{\plotpoint}}
\multiput(311,169)(14.676,-14.676){0}{\usebox{\plotpoint}}
\multiput(316,164)(17.270,-11.513){0}{\usebox{\plotpoint}}
\put(326.13,157.24){\usebox{\plotpoint}}
\multiput(328,156)(17.798,-10.679){0}{\usebox{\plotpoint}}
\multiput(333,153)(18.564,-9.282){0}{\usebox{\plotpoint}}
\multiput(339,150)(19.271,-7.708){0}{\usebox{\plotpoint}}
\put(344.56,147.81){\usebox{\plotpoint}}
\multiput(350,146)(19.690,-6.563){0}{\usebox{\plotpoint}}
\multiput(356,144)(19.271,-7.708){0}{\usebox{\plotpoint}}
\put(364.14,140.95){\usebox{\plotpoint}}
\multiput(367,140)(20.473,-3.412){0}{\usebox{\plotpoint}}
\multiput(373,139)(19.271,-7.708){0}{\usebox{\plotpoint}}
\multiput(378,137)(20.473,-3.412){0}{\usebox{\plotpoint}}
\put(384.19,135.97){\usebox{\plotpoint}}
\multiput(390,135)(20.595,-2.574){3}{\usebox{\plotpoint}}
\multiput(446,128)(20.705,-1.453){2}{\usebox{\plotpoint}}
\multiput(503,124)(20.742,-0.741){3}{\usebox{\plotpoint}}
\multiput(559,122)(20.743,-0.728){3}{\usebox{\plotpoint}}
\multiput(616,120)(20.752,-0.364){2}{\usebox{\plotpoint}}
\multiput(673,119)(20.752,-0.371){3}{\usebox{\plotpoint}}
\multiput(729,118)(20.752,-0.364){3}{\usebox{\plotpoint}}
\multiput(786,117)(20.756,0.000){3}{\usebox{\plotpoint}}
\multiput(842,117)(20.752,-0.364){2}{\usebox{\plotpoint}}
\put(902.31,116.00){\usebox{\plotpoint}}
\put(910,116){\usebox{\plotpoint}}
\sbox{\plotpoint}{\rule[-0.200pt]{0.400pt}{0.400pt}}%
\multiput(222.60,620.87)(0.468,-10.717){5}{\rule{0.113pt}{7.500pt}}
\multiput(221.17,636.43)(4.000,-58.433){2}{\rule{0.400pt}{3.750pt}}
\multiput(226.59,550.35)(0.477,-9.059){7}{\rule{0.115pt}{6.660pt}}
\multiput(225.17,564.18)(5.000,-68.177){2}{\rule{0.400pt}{3.330pt}}
\multiput(231.59,477.32)(0.482,-5.915){9}{\rule{0.116pt}{4.500pt}}
\multiput(230.17,486.66)(6.000,-56.660){2}{\rule{0.400pt}{2.250pt}}
\multiput(237.59,414.64)(0.482,-4.830){9}{\rule{0.116pt}{3.700pt}}
\multiput(236.17,422.32)(6.000,-46.320){2}{\rule{0.400pt}{1.850pt}}
\multiput(243.59,361.31)(0.477,-4.718){7}{\rule{0.115pt}{3.540pt}}
\multiput(242.17,368.65)(5.000,-35.653){2}{\rule{0.400pt}{1.770pt}}
\multiput(248.59,323.18)(0.482,-3.022){9}{\rule{0.116pt}{2.367pt}}
\multiput(247.17,328.09)(6.000,-29.088){2}{\rule{0.400pt}{1.183pt}}
\multiput(254.59,290.56)(0.482,-2.570){9}{\rule{0.116pt}{2.033pt}}
\multiput(253.17,294.78)(6.000,-24.780){2}{\rule{0.400pt}{1.017pt}}
\multiput(260.59,261.95)(0.477,-2.491){7}{\rule{0.115pt}{1.940pt}}
\multiput(259.17,265.97)(5.000,-18.973){2}{\rule{0.400pt}{0.970pt}}
\multiput(265.59,241.33)(0.482,-1.666){9}{\rule{0.116pt}{1.367pt}}
\multiput(264.17,244.16)(6.000,-16.163){2}{\rule{0.400pt}{0.683pt}}
\multiput(271.59,223.16)(0.482,-1.395){9}{\rule{0.116pt}{1.167pt}}
\multiput(270.17,225.58)(6.000,-13.579){2}{\rule{0.400pt}{0.583pt}}
\put(277,212){\usebox{\plotpoint}}
\put(277,212){\usebox{\plotpoint}}
\multiput(277.59,207.27)(0.477,-1.378){7}{\rule{0.115pt}{1.140pt}}
\multiput(276.17,209.63)(5.000,-10.634){2}{\rule{0.400pt}{0.570pt}}
\multiput(282.59,195.82)(0.482,-0.852){9}{\rule{0.116pt}{0.767pt}}
\multiput(281.17,197.41)(6.000,-8.409){2}{\rule{0.400pt}{0.383pt}}
\multiput(288.59,186.09)(0.482,-0.762){9}{\rule{0.116pt}{0.700pt}}
\multiput(287.17,187.55)(6.000,-7.547){2}{\rule{0.400pt}{0.350pt}}
\multiput(294.59,177.26)(0.477,-0.710){7}{\rule{0.115pt}{0.660pt}}
\multiput(293.17,178.63)(5.000,-5.630){2}{\rule{0.400pt}{0.330pt}}
\multiput(299.59,170.65)(0.482,-0.581){9}{\rule{0.116pt}{0.567pt}}
\multiput(298.17,171.82)(6.000,-5.824){2}{\rule{0.400pt}{0.283pt}}
\multiput(305.00,164.93)(0.599,-0.477){7}{\rule{0.580pt}{0.115pt}}
\multiput(305.00,165.17)(4.796,-5.000){2}{\rule{0.290pt}{0.400pt}}
\multiput(311.00,159.94)(0.627,-0.468){5}{\rule{0.600pt}{0.113pt}}
\multiput(311.00,160.17)(3.755,-4.000){2}{\rule{0.300pt}{0.400pt}}
\multiput(316.00,155.94)(0.774,-0.468){5}{\rule{0.700pt}{0.113pt}}
\multiput(316.00,156.17)(4.547,-4.000){2}{\rule{0.350pt}{0.400pt}}
\multiput(322.00,151.95)(1.132,-0.447){3}{\rule{0.900pt}{0.108pt}}
\multiput(322.00,152.17)(4.132,-3.000){2}{\rule{0.450pt}{0.400pt}}
\multiput(328.00,148.95)(0.909,-0.447){3}{\rule{0.767pt}{0.108pt}}
\multiput(328.00,149.17)(3.409,-3.000){2}{\rule{0.383pt}{0.400pt}}
\multiput(333.00,145.95)(1.132,-0.447){3}{\rule{0.900pt}{0.108pt}}
\multiput(333.00,146.17)(4.132,-3.000){2}{\rule{0.450pt}{0.400pt}}
\put(339,142.17){\rule{1.100pt}{0.400pt}}
\multiput(339.00,143.17)(2.717,-2.000){2}{\rule{0.550pt}{0.400pt}}
\put(344,140.67){\rule{1.445pt}{0.400pt}}
\multiput(344.00,141.17)(3.000,-1.000){2}{\rule{0.723pt}{0.400pt}}
\put(350,139.17){\rule{1.300pt}{0.400pt}}
\multiput(350.00,140.17)(3.302,-2.000){2}{\rule{0.650pt}{0.400pt}}
\put(356,137.67){\rule{1.204pt}{0.400pt}}
\multiput(356.00,138.17)(2.500,-1.000){2}{\rule{0.602pt}{0.400pt}}
\put(361,136.17){\rule{1.300pt}{0.400pt}}
\multiput(361.00,137.17)(3.302,-2.000){2}{\rule{0.650pt}{0.400pt}}
\put(367,134.67){\rule{1.445pt}{0.400pt}}
\multiput(367.00,135.17)(3.000,-1.000){2}{\rule{0.723pt}{0.400pt}}
\put(373,133.67){\rule{1.204pt}{0.400pt}}
\multiput(373.00,134.17)(2.500,-1.000){2}{\rule{0.602pt}{0.400pt}}
\put(378,132.67){\rule{1.445pt}{0.400pt}}
\multiput(378.00,133.17)(3.000,-1.000){2}{\rule{0.723pt}{0.400pt}}
\put(384,131.67){\rule{1.445pt}{0.400pt}}
\multiput(384.00,132.17)(3.000,-1.000){2}{\rule{0.723pt}{0.400pt}}
\multiput(390.00,130.93)(6.165,-0.477){7}{\rule{4.580pt}{0.115pt}}
\multiput(390.00,131.17)(46.494,-5.000){2}{\rule{2.290pt}{0.400pt}}
\multiput(446.00,125.93)(6.276,-0.477){7}{\rule{4.660pt}{0.115pt}}
\multiput(446.00,126.17)(47.328,-5.000){2}{\rule{2.330pt}{0.400pt}}
\put(503,120.17){\rule{11.300pt}{0.400pt}}
\multiput(503.00,121.17)(32.546,-2.000){2}{\rule{5.650pt}{0.400pt}}
\put(559,118.67){\rule{13.731pt}{0.400pt}}
\multiput(559.00,119.17)(28.500,-1.000){2}{\rule{6.866pt}{0.400pt}}
\put(616,117.67){\rule{13.731pt}{0.400pt}}
\multiput(616.00,118.17)(28.500,-1.000){2}{\rule{6.866pt}{0.400pt}}
\put(673,116.67){\rule{13.490pt}{0.400pt}}
\multiput(673.00,117.17)(28.000,-1.000){2}{\rule{6.745pt}{0.400pt}}
\put(786,115.67){\rule{13.490pt}{0.400pt}}
\multiput(786.00,116.17)(28.000,-1.000){2}{\rule{6.745pt}{0.400pt}}
\put(729.0,117.0){\rule[-0.200pt]{13.731pt}{0.400pt}}
\put(842.0,116.0){\rule[-0.200pt]{16.381pt}{0.400pt}}
\end{picture}

\caption{Comparison between the simulation result for $c({\overline
t})$ (solid line) and the value given by Eqn.  (\ref{acvf02}) (dotted
line) using the simulation result for the environment velocity
correlation function $C({\overline t})$. Parameters: $s=2.82,
\Omega=25, \mu=10.0$ }
\label{fig.check} 
\end{center} 
\end{figure}

In order to derive Eqn. (\ref{acvf02}) we made the approximation that
the density field is approximately constant. We check now that this
assumption was reasonable by computing in a simulation the environment
velocity correlation $C({\overline t})$, evaluating numerically the
integral term in Eqn. (\ref{acvf02}), and adding up the first
exponential term in Eqn. (\ref{acvf02}). The result is the dotted line
in Fig. \ref{fig.check}. Also shown is the result for the velocity
correlation function $c({\overline t})$ obtained directly from the
simulation (solid line). We see that both results are in quite good
agreement giving confidence on Eqn. (\ref{acvf02}) as a sounded
starting point for theoretical analysis. Similar good agreement is
obtained for all the values of $\Omega$ studied ($\Omega=0.5,8.3,25$)

\begin{figure}[ht]
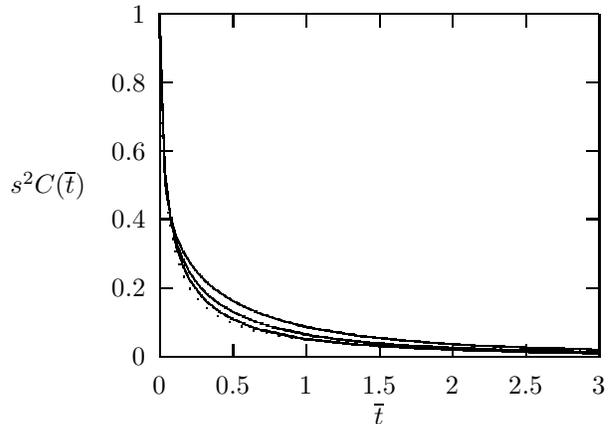

\begin{center}
\setlength{\unitlength}{0.240900pt}
\ifx\plotpoint\undefined\newsavebox{\plotpoint}\fi


\label{fig2}
\caption{Simulation results for the 
velocity correlation function $s^2 C({\overline t})$ for a
fixed value of $\Omega=25$ and different values of the overlapping 
coefficient $s=1.5,2.18,3.2$ (upper curve correspond to lower $s$). 
Dotted line is the hydrodynamic prediction (\ref{hp3}). As the overlapping
increases, the simulation results converge towards the theoretical prediction.}
\end{center}
\end{figure}

For a given value of $\Omega=25$ we plot in Fig. \ref{fig2} the value
of $s^2 C({\overline t})$, for different values of
$s$. According to the hydrodynamic prediction (\ref{hp3}) this curve
should be independent of the overlapping coefficient $s$ (dotted curve
in Fig. \ref{fig2}). We observe that the simulation results converge
towards the theoretical prediction only when the overlapping
coefficient is sufficiently large. This is expected from the fact that
hydrodynamic behavior appears only on length scales that involve a
relatively large number of particles.

We have investigated also the effect of the finite system size. We
plot in Fig. \ref{fig2} the environment velocity correlation function
for different system sizes, while keeping the rest of parameters
constant ($s=2.82$, $\Omega= 25$). We observe a large discrepancy
between the hydrodynamic prediction and the simulation results when
the box is small. This discrepancy appears at large times and is
reduced when the system size increases. This effect can be understood
as an self-interaction through periodic boundary conditions.  According
to Onsager hypothesis for the regression of fluctuations, the
correlation of the velocity fluctuations decays in essentially the
same way as a localized macroscopic velocity perturbation. A
perturbation in the velocity field in the form of an impulse at the
origin decays through the formation of a vortex ring whose center
evolves diffusively as $\sqrt{t}$ \cite{esp95c}. In a system with periodic
boundary conditions the vortex ring is reintroduced into the system
producing an effective slowing down of the velocity field at the
origin. Obviously, the larger is the system size, the latter the
effect appears and also the smaller is the amplitude of the effect.

\begin{figure}[ht]
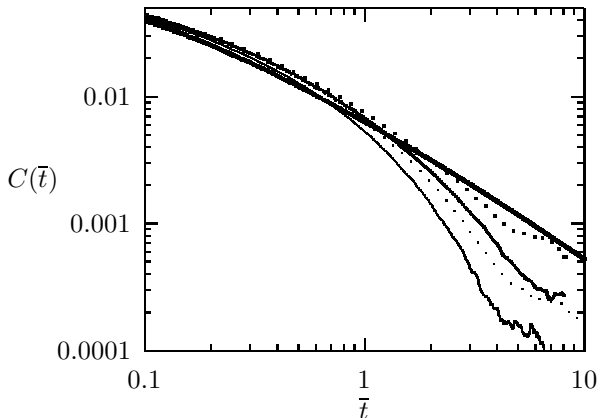

\begin{center}
\setlength{\unitlength}{0.240900pt}
\ifx\plotpoint\undefined\newsavebox{\plotpoint}\fi


\label{figbox}
\caption{Simulation results for the velocity correlation function
$ C({\overline t})$ for fixed value of $s=2.82,\Omega=25$
and different values of the box size $\mu=7.9,11.2,15.8,35.5$
(lower curve correspond to lower $\mu$). Thicker line is the
hydrodynamic prediction (\ref{hp3}).}
\end{center}
\end{figure}

\begin{figure}[ht]
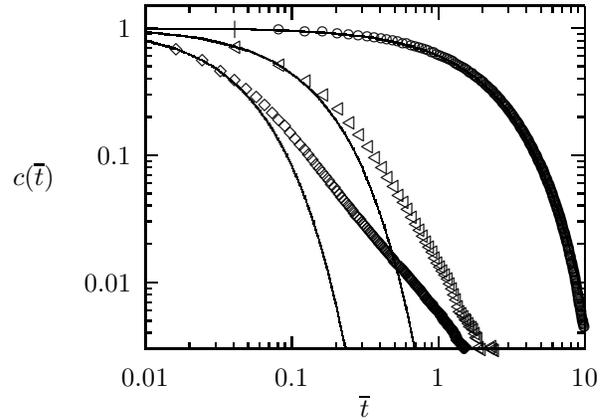

\begin{center}
\setlength{\unitlength}{0.240900pt}
\ifx\plotpoint\undefined\newsavebox{\plotpoint}\fi

\label{fig.omega}
\caption{Simulation results for the velocity correlation function
$c({\overline t})$ for fixed value of $s=2.82,\mu=10.0$
and different values of $\Omega=0.5$ (circles),
$\Omega=8.3$ (triangles) and $\Omega=25$ (diamonds). Solid lines 
are the term $\exp\{-\Omega {\overline t}\}$ in Eqn. (\ref{acvf02}).}
\end{center}
\end{figure}

Next, we study the effect of $\Omega$ on the velocity
autocorrelation function of the DPD particles. In
Fig. \ref{fig.omega} we show the velocity correlation function for
given $s=2.82, \mu=10.0$ and three different values of
$\Omega=0.5,8.3,25$. We also plot  the corresponding exponential
terms in Eqn. (\ref{acvf02}). For small $\Omega$ (small friction or high
temperature) the decay of the vaf is very accurately given by the
exponential term. As long as $\Omega$ increases, discrepancies from
the exponential behavior are observed. This discrepancies are due to
the effect of the collective term in Eqn. (\ref{acvf02}).

\begin{figure}[ht]
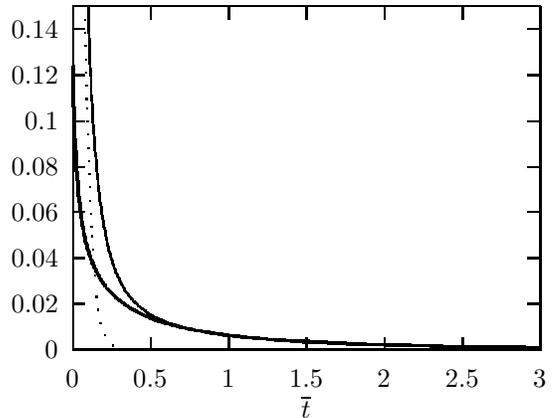

\begin{center}
\setlength{\unitlength}{0.240900pt}
\ifx\plotpoint\undefined\newsavebox{\plotpoint}\fi
\sbox{\plotpoint}{\rule[-0.200pt]{0.400pt}{0.400pt}}%


\label{fig.sim}
\caption{Simulation results for the velocity correlation function
$c({\overline t})$ (thin line) and the environment velocity
correlation function $C({\overline t})$ (thick
line). Also shown is the term $\exp\{-\Omega {\overline t}\}$ in
Eqn. (\ref{acvf02}). Here, $s=2.82,\mu=10.0, \Omega=25$.}
\end{center} 
\end{figure}

In Fig. \ref{fig.sim} it is shown that for large $\Omega$, at large times the 
environment and particle velocities coincide
according to Eqn. (\ref{highomeg}). 

\section{Summary and discussion}

We have presented a theory for Dissipative Particle Dynamics that
allows to understand the different dynamical regimes displayed by the
model. The theory is based on the physical picture that DPD particles
behave like Brownian particles in a non-equilibrium environment due to
the rest of DPD particles. An explicit expression for the velocity
autocorrelation function is derived in which the Brownian
exponential behavior is corrected by the presence of collective
effects. By using dimensionless variables it is possible to asses the
range of parameters for which the collective effects are important.
Three dimensionless parameters appear in the model, $s,\Omega,\mu$,
and they characterize the different dynamical regimes in the system.
The relevance of precisely these dimensionless groups is motivated by
the theory presented (in Ref. \cite{pag98} other dimensionless groups are
introduced). 

Two dynamical regimes are identified, the mean field regime and the
collective behavior regime. The transition between both is governed
essentially by the dimensionless friction $\Omega$ with important
effects of $s$ and $\mu$. Mean field behavior appears for small
friction $\Omega$ or large overlapping $s$. This is physically
reasonable: For small friction, the dynamics of the environment of a
given particle hardly affects the behavior of this particle. For large
overlapping the non-equilibrium collective effects are smeared out
over large regions. The mean field approximation is closely related to
the molecular chaos assumption made in kinetic theory. Actually, it is
possible to compute the diffusion coefficient of the DPD particles by
using the mean field approximation for the vaf into the usual
Green-Kubo formula for the diffusion coefficient \cite{gro97}. The result is
precisely the prediction for the diffusion coefficient given by
kinetic theory \cite{mar97} (with due account of the normalization
given in Eqn. (\ref{norm})), i.e.

\begin{equation}
D = \int_0^\infty 
\frac{1}{d}\langle {\bf v}_i(t)\!\cdot\!{\bf v}_i(0)\rangle
= \frac{d}{\gamma}\frac{k_BT}{m}.
\label{dif}
\end{equation}
When the dimensionless friction $\Omega$ is high, the vaf does not
decay in an exponential way because it is dominated by the collective
dynamics. We have presented a theoretical prediction for the
collective part of the vaf by assuming that the correlation function
of the environment velocity reflects an underlying hydrodynamic
behavior. Such a behavior is expected (and observed in the
simulations) when the overlapping coefficient $s$ is large enough. In
this case, the collective effect is small but well described by
hydrodynamics.

The fact that hydrodynamics governs the dynamics of the velocity of
the particles has two important consequences. The first one is that
finite size effects are important. Hydrodynamic self-interaction
through the periodic boundary conditions exists and large box sizes
must be considered in order to render this effect small. This
self-interaction occurs not only through the sound mode but also
through the shear mode \cite{web}. The second consequence is the appearance of
the celebrated long-time tails in the vaf. These algebraic tails
occur at very long times (for which the vaf has decayed to a factor
$10^{-3}$ from its original value) and are difficult to observe in our
simulations due to the statistical noise. Nevertheless, we have
provided sufficient numerical evidence for the hydrodynamic behavior
at smaller times and we expect the presence of long time tails at
large times for sufficiently large system sizes. The $t^{-1}$
dependence of the vaf, when introduced into the Green-Kubo formula
(\ref{dif}) leads to a logarithmically divergent diffusion
coefficient. It is apparent that this ``small'' divergence will be
hardly observable in any simulation with a finite box size.

In the regimes for which collective effects are important, we expect
that deviations from the predictions of kinetic theory occur not only
for the diffusion coefficient, but also for the rest of the transport
coefficients of the DPD fluid \cite{pag98}. It is actually possible to
use the theory presented in this paper in order to compute the
transport coefficients for the DPD fluid expressed in the form of
Green-Kubo formulae \cite{esp95b}. One obtains then a set of recursive
equations in which the transport coefficients are expressed in terms
of the transport coefficients used in the hydrodynamic assumption. We
do not follow this rather cumbersome numerical procedure here because
it does not provide any new physical insight.

Dissipative Particle Dynamics is designed to simulate hydrodynamic
problems. Actually, one would like to have the DPD particles
moving in such a way that they follow accurately the flow field
intended to be modeled. We see that this will happen whenever the
friction is sufficiently large (in this case the velocity of a DPD
particle is slaved by its environment) and the overlapping is
sufficiently large (in such a way that the environment velocity moves
hydrodynamically). In this regime, the dynamics of the particles is
mainly collective and kinetic theory gives inaccurate values for the
transport coefficients \cite{pag98}.

\section*{Acknowledgments}
We are grateful to M.H. Ernst, P. Warren, M. Ripoll, I. Z\'u\~niga,
and M. Revenga for stimulating discussions.

\section*{Appendix I}

We derive the equilibrium value for the environment velocity
correlation function given in Eqn. (\ref{evt0}). From the definition
of ${\bf V}_i$ in Eqn. (\ref{vhidro}) we calculate the mean value by
using the canonical ensemble

\begin{eqnarray}
\frac{1}{d}\langle {\bf V}_i(0)\!\cdot\! {\bf V}_i(0) \rangle 
&=&d\langle(\sum_{j\neq i}\omega_{ij}{\bf e}_{ij}{\bf e}_{ij}
\!\cdot\!{\bf v}_j)(\sum_{k\neq i}\omega_{ik}{\bf e}_{ik}{\bf e}_{ik}
\!\cdot\!{\bf v}_k) \rangle \nonumber\\
&=& \frac{d k_B T (N-1)}{m}\frac{\int d{\bf r }_id{\bf r }_j
 \omega(r_{ij})}{\int d{\bf r }_id{\bf r }_j} \nonumber\\
&=&\frac{d k_B T}{ m} \frac{N-1}{N}\frac{3}{2 \pi r_c^2 n} 
\nonumber\\
&\approx&\frac{d k_B T}{ m} \frac{3}{2 \pi s^2},
\label{ve0}
\end{eqnarray}
for large  $N$  (number of DPD particles in the system) and  
$n$ the number density.

\section*{Appendix II}

On simple symmetry grounds, the tensor ${\bf \omega}({\bf k})$ has the form

\begin{equation}
{\bf \omega}({\bf k}) = \int d{\bf r}\omega(r){\bf  \hat{r}}{\bf \hat{r}} 
\exp\{-i {\bf k }{\bf r}\} = a(kr_c){\bf 1}+ b(kr_c){\bf  \hat{k}}{\bf \hat{k}}
\end{equation}
In order to calculate the functions $a(kr_c)$ and $b(kr_c)$ we double
contract ${\bf \omega}({\bf k})$ with the dyadic ${\bf\hat{k}}{\bf
\hat{k}}$ and also take its trace. This leads to

\begin{eqnarray}
{\bf  \hat{k}^T}{\bf \omega}(k){\bf \hat{k}} 
&=&
a(kr_c)+b(kr_c)
\nonumber\\
&=& 2 \pi \int_0^\infty dr \omega(r) \left(\frac{J_{1}(kr)}{k}- r J_2(kr)\right),
\nonumber\\
{\rm tr}({\bf \omega}(k)) &=& 2 a(kr_c)+ b(kr_c) 
\nonumber\\
&=& 2 \pi \int_0^\infty r \omega(r)J_{o}(kr).
\end{eqnarray}
The integrals are given in terms of generalized hypergeometric functions 
$_p F_q\{{\bf a},{\bf b},z\}$  \cite{mathematica} and Bessel functions
\begin{equation}
_p F_q\{{\bf a},{\bf b},z\}= 
\sum_{k=0}^{\infty}\frac{(a_1)_k...(a_p)_k}{(b_1)_k...(b_q)_k}\frac{z^k}{k! }
\end{equation}
where $(m)_k=m(m+1)...(m+k-1)$. More precisely,
\begin{eqnarray}
a(kr_c)+b(kr_c) &=&
-\frac{1}{n} {_1 F_2 }\{( \frac{3}{2} ),(2,\frac{5}{2}),-\frac{(k r_c)^2}{4}\}
\nonumber\\
&-& \frac{6}{n (k r_c)^2}+ \frac{6}{n (k r_c)^2}J_{0}(k r_c) 
+ \frac{6}{n k r_c} J_{1}(k r_c)
\nonumber\\
&+&\frac{6 (k r_c)^2}{n} 
{_1 F_2}\{ (\frac{5}{2}),(3,\frac{7}{2}),-\frac{(k r_c)^2}{4}\}
\nonumber\\
2 a(kr_c)+ b(kr_c) &=& \frac{6 J_{1} (kr_c)}{k r_c n}
-\frac{2}{n} {_1 F_2 }\{(\frac{3}{2}),(1,\frac{5}{2}),-\frac{(k r_c)^2}{4}\}
\end{eqnarray}
The solution of this system of two equations provides 
the values for $a(kr_c)$ and $b(kr_c)$.

\end{document}